\documentclass[showpacs,aps,preprintnumbers,nofootinbib]{revtex4}
\usepackage{graphicx}

\newcommand{\be}{\begin{equation}}
\newcommand{\ee}{\end{equation}}
\newcommand{\bel}[1]{\be\label{#1}}

\newcommand{\ber}{\begin{eqnarray}}
\newcommand{\eer}{\end{eqnarray}}

\newcommand{\psib}{\overline{\psi}}

\newcommand{\hsp}{\hspace*{1pt}}

\newcommand{\bg}{B^{1/4}}
\newcommand{\rb}{\rho_B}
\newcommand{\dn}{{\rm fm}^{-3}}

\begin{document}
\title {Deconfinement Phase Transition in Compact Stars : 
Maxwell vs. Gibbs Construction of the Mixed Phase}

\author{Abhijit Bhattacharyya$^{1,2}$, Igor N. Mishustin$^{2,3}$ 
and Walter Greiner$^{2}$}

\affiliation{${}^1${\it Department of Physics, University of Calcutta, 
92, A. P. C. Road, Kolkata 700 009, India} \\
${}^2${\it Frankfurt Institute for Advanced Studies, 
J.W.~Goethe--Universit\"at, Ruth-Moufang-Strasse 1,}
{\it D--60438~Frankfurt~am~Main,~Germany} \\
${}^3${\it The Kurchatov Institute, Russian Research
Center, 123182 Moscow, Russia}}

\begin{abstract}  
We study different realisations of the first order deconfinement 
phase transition inside a 
compact star by comparing the Gibbs and Maxwell construction for 
the mixed phase. The hadronic sector is described within the 
relativistic mean field model including hyperons. The quark 
sector is described by the MIT Bag model. We find that these two 
realisations lead 
to very different star properties, in particular, the composition 
of the stellar matter. We also find that for the Maxwell 
construction there is a sharp discontinuity in the baryon density 
and the electron chemical 
potential. We argue that a sharp jump in the elctron chemical 
potential should lead to the redistribution of electrons and formation 
of strong electric fields around the discontinuity surface. 
\end{abstract}

\maketitle

\section{Introduction}

The heavy ion experiments at RHIC, LHC and FAIR are designed to 
study strongly interacting matter under extreme conditions of 
high temperature and/or high baryon density. These experiments 
are also expected to shed light on the properties of a new 
phase of strongly interacting matter, the Quark Gluon Plasma 
(QGP) and on the nature of the deconfinement phase transition. 
On the other hand, compact stars, due to their large central 
densities, serve as a natural laboratory to study the 
properties of the strongly interacting matter at high densities 
and small temperature, in particular, the possibility of a 
deconfining phase transition. 

At finite temperature and zero baryon density, numerical studies 
on lattice are believed to provide reliable results for the physics 
of the deconfinement transition \cite{karsch}. 
In this case the lattice calculations predict
that the deconfinement happens via a smooth crossover transition 
\cite{aoki} at a temperature $\sim$ 170 - 200 MeV \cite{tc}.
However, the studies at finite baryon densities on the lattice are very 
difficult. Some progress has been made in recent
years in extending calculations to finite quark chemical potentials but 
they do not provide reliable results yet \cite{fodor,philipsen}. 
There is an indication of a critical point at a rather small
quark chemical potential $\mu _q \approx$ 100 MeV, with a first order 
transition for larger $\mu_q$ \cite{fodor}. 

The most recent interest in the study of first order deconfining 
phase transition is related to the nature of the 
mixed phase (MP) \cite{glenphysrep}. Especially, the role 
of the screened Colulomb 
potential and interface effects in the MP were studied by several 
authors (see e.g. refs. \cite{pasta,maruyama,maruyama1,vos,chiba}). 
A detailed study employing the Wigner Seitz cell approach \cite{pasta} 
suggests that the MP behaves more in accordance with that for Maxwell 
Construction (MC) rather than the Gibbs Construction (GC). 
Earlier similar questions were addressed in connection with 
the "pasta" phases associated with the liquid-gas phase transition 
\cite{raven}.

In this work we present a comparative study of compact stars with the 
MC and GC of the mixed phase. For the
hadronic phase we use a Relativistic Mean Field (RMF) model of the  
Walecka type \cite{serot}. Besides nucleons, this
model contains hyperons as well as hyperon-hyperon interaction. For 
the quark sector we employ the MIT Bag model \cite{cho74}  which has 
been used previously for the description of strange quark stars (see a 
recent review \cite{bombaci}). We construct 
the MP for the deconfinement phase transition and obtain the equations 
of state (EOSs) for the both cases. 
The next section is devoted to the description of these models and 
corresponding EOSs. In 
section 3 we use these EOSs to describe properties of the compact 
stars. In section 4 we study the jump in the electron chemical 
potential at the interface between the two phases and estimate 
an induced electric field. In the last section we summarise our 
results.

\section{Properties of matter in compact stars} 

\subsection{Hadron phase}

At low densities the relevant degrees of freedom are hadrons. 
To describe the hadronic phase we use a non-linear version of the 
RMF model. In this model the baryons interact with mean  
meson fields. The variant that we use here is known as the TM1 
model \cite{toki}.

The Lagrangian density for the TM1 model including both nucleons and 
hyperons is written as~\cite{toki,scha96}
\ber
{\cal L}&=&\sum_B \psib_B\left(i\rlap /{\partial} -m_B \right) \psi_B + {1 \over 2} 
\partial^\mu \sigma \partial_\mu \sigma  
-\frac{1}{2} m_\sigma^2 \sigma^2
-\frac{b}{3}\sigma^3-\frac{c}{4}\sigma^4 
-{1 \over 4} \omega^{\mu\nu} \omega_{\mu\nu}+  
\frac{1}{2} m_\omega^2 \hsp\omega^\mu\omega_\mu \nonumber\\
&+& \frac{d}{4} (\omega_\mu \omega^\mu)^2 
-{1 \over 4} {\vec\rho}^{\,\mu\nu} {\vec\rho}_{\mu\nu}+ 
{1 \over 2} m_\rho^2 {\vec\rho}^{\mu} {\vec\rho}_{\hsp\mu}
+\sum_B \psib_B \left(g_{\sigma B} \sigma + 
g_{\omega B}\omega^\mu\gamma_\mu 
+ g_\rho{\vec\rho}^{\,\mu}\gamma_\mu {\vec\tau}_B\right)\psi_B\,,
\label{lagr}
\eer
where the sum runs over all the baryons $B$=$p, n, \Lambda,\Sigma^{0,\pm}, 
\Xi^{0,-}$. In the above Lagrangian $\sigma, \omega$ and
${\vec\rho}$ are respectively the iso-scalar scalar $\sigma$, the 
iso-scalar vector $\omega$ and the isovector vector $\rho$ meson fields. 
In eq.~(\ref{lagr}) $\omega^{\mu\nu}$ and ${\vec\rho}^{\,\mu\nu}$ denote,
respectively, the field tensors for the $\omega$ and 
$\rho$ meson fields.

This model is good enough to describe nucleonic matter and the nuclear 
saturation point. But it is insufficient for the hyperonic matter,
because the model does not reproduce the observed strong 
$\Lambda \Lambda$ 
attraction. This defect can be remedied by adding two new meson
fields with hidden strangeness,  namely, the iso-scalar scalar $\sigma^*$ and 
the iso-vector vector $\phi$, which couple to hyperons only~\cite{scha96}.
These fields can be identified with the $f_0\hsp (975)$ and $\phi\hsp (1020)$ 
mesons. The corresponding Lagrangian is given by
\ber
{\cal L}^{YY}={1 \over 2} \left(\partial^\mu \sigma^* \partial_\mu \sigma^* 
- m^2_{\sigma^*} \sigma^{*2}\right)  - {1 \over 4} \phi^{\mu\nu}
\phi_{\mu\nu}+{1 \over 2} m_\phi^2 \phi^\mu \phi_\mu 
+ \sum_Y \psib_Y \left(g_{\sigma^* Y} \sigma^* 
+ g_{\phi Y} \phi^\mu\gamma_\mu \right)\psi_Y \nonumber
\eer
where index $Y$ runs over hyperons only. 

For a complete description of the beta equilibrated cold matter the model 
should include leptons; namely electrons and muons. The leptonic part 
of the lagrangian is 
\be
{\cal L}^l = \sum_{l=e^-,\mu^-} \psib_l\left(i\rlap / {\partial} -m_B \right) \psi_l 
\ee
The Lagrangian density of the complete model, which we call TM1YY, is 
written as 
\be
{\cal L}^{TM1YY} = {\cal L} + {\cal L}^{YY} + {\cal L}^l
\ee

The nucleon coupling constants are chosen from the fit of the finite nuclei
properties. The vector coupling constants of the hyperons are chosen according
to the SU(6) symmetry and the hyperonic scalar coupling constants are chosen to
reproduce the measured values of the corresponding optical potentials.
Below we use the set of model parameters obtained in 
ref.~\cite{scha96}.

We calculate the energy density and pressure for the TM1YY model in the 
mean field approximation. They are given by the following expressions

\ber
\epsilon^{H} &=& \frac{1}{2}m_\sigma^2 \sigma^2
+ \frac{b}{3}\sigma^3 + \frac{c}{4}\sigma^4
+ \frac{1}{2}m_{\sigma^*}^2 {\sigma^*}^2
+ \frac{1}{2}m_\omega^2 \omega_0^2 + \frac{3d}{4} \omega_0^4
\cr && {}
+ \frac{1}{2}m_\rho^2 \rho_{0,0}^2
+ \frac{1}{2}m_\phi^2 \phi_0^2
+ \sum_{B} \frac{\nu_B}{2\pi^2} \int_0^{k_F^B} dk\hsp k^2
\sqrt{k^2 + {m^*_B}^2}~, \\ \cr
P^{H} &=& - \frac{1}{2}m_\sigma^2 \sigma^2
- \frac{b}{3}\sigma^3 - \frac{c}{4}\sigma^4
- \frac{1}{2}m_{\sigma^*}^2 {\sigma^*}^2
+ \frac{1}{2}m_\omega^2 \omega_0^2 + \frac{d}{4} \omega_0^4
\cr && {}
+ \frac{1}{2}m_\rho^2 \rho_{0,0}^2
+ \frac{1}{2}m_\phi^2 \phi_0^2
+ \sum_{B} \frac{\nu_B}{6\pi^2} \int_0^{k_F^B} dk
\frac{k^4}{\sqrt{k^2 + {m^*_B}^2}}\,,
\eer
where $m^*_B=m_B-g_{\sigma B}\sigma-g_{\sigma^* B}\sigma^*$ is the effective 
mass, $\nu_B$ is the degeneracy factor and $k_F^B=\sqrt{\mu_B^2-m^{*2}_B}$ 
is the Fermi momentum  of the baryon species $B$.

\subsection{Quark Phase}

At higher densities baryons begin to overlap and loose their 
individuality. In order to describe 
the medium the quark degrees of freedom need to be included. 
The density inside a compact star is high enough to encompass these 
degrees of freedom. In order to describe the quark phase we adopt the 
simple MIT Bag model~\cite{cho74}, with three flavours (u, d and s). 
We also add electrons and muons to describe the beta equilibrated 
matter as in the case of the hadronic phase. 
For the bag model the energy density and pressure 
can be written as 
\begin{eqnarray}
\epsilon^Q &=& \sum_{f=u,d,s} 
\frac{\nu_f}{2 \pi^2} \int_0^{k_F^f} dk k^2\sqrt{m_f^2 + k^2}+ 
B\,,\label{edec}\\ 
P^Q &=& \sum_{f=u,d,s} \frac{\nu_f}{6\pi^2} 
\int_0^{k_F^f} dk \frac{k^4}{\sqrt{m_f^2 + k^2}}- B\,, 
\label{pdec}
\end{eqnarray}
where $k_F^f=\sqrt{\mu_f^2-m_f^2}$ is the Fermi momentum of quarks  with 
flavor~$f$.  For each flavor we choose the degeneracy factor  
$\nu_f = 2\hsp ({\rm spin}) \times 3\hsp ({\rm color}) = 6$ and take the 
following values of quark masses: $m_u=5$ MeV, $m_d=10$ MeV and $m_s=150$~MeV.

It is worth noting that because of the negative vacuum pressure 
($-B$ in eq.(7)) the Bag model EOS always has a zero pressure 
at a finite baryon density, $\rho_B^* \sim B^{3/4}$. By this reason 
the equilibrium configurations of the strange quark matter (SQM) may 
exist even without gravity \cite{farhi}. They should have a sharp 
boundary with the density jump from $\rho_B^*$ to zero. 
The EOS derived from the NJL model has the similar property \cite{hanauske}.

\subsection{Construction of the mixed phase and EOS}

We are studying the electrically neutral stellar matter in beta 
equilibrium. Under such conditions the chemical potential of a 
particle species {\it i} can be written as 
\bel{chem}
\mu_i = B_i\hsp\mu_B + Q_i\hsp\mu_Q
\ee
where $B_i$ is the baryon number of the species $i$\,,
$Q_i$ denotes its charge in units of the electron charge, 
$\mu_B$ and $\mu_Q$ are the baryonic and electric chemical potentials,
respectively. Here we assume that neutrinos can freely escape
from the star. The above equation signifies that only those reactions 
are allowed which conserve charge and baryon number, however strangeness 
is not conserved. Two independent chemical
potentials, $\mu_B$ and $\mu_Q$, are found by fixing the baryon 
and electric charge densities:
\begin{equation}
\rho_B=\sum_i B_i \rho_i\,,~~~~\rho_Q=\sum_i Q_i \rho_i\,,
\end{equation}
where $\rho_i$ is the number density of the particle species $i$\,.
We require electrical neutrality of a star 
on a macroscopic scale, {\it i.e.}~\mbox{${\bar \rho}_Q=0$}. 
According to Eq.~(\ref{chem}), 
the baryon chemical potential
$\mu_B$ equals the neutron chemical potential $\mu_n$\, and
$\mu_Q$ is equal to the electron chemical potential $\mu_e$.
At given $\mu_B$ and $\mu_Q$, the quark chemical potentials
are found by using the formulae
$\mu_u=(\mu_B-2\hsp\mu_Q)/3$ and $\mu_d=\mu_s =(\mu_B+\mu_Q)/3$\,.

As indicated by the model calculations (see e.g. ref.\cite{plb}) 
the deconfinement phase 
transition, at high densities, is of first order in nature. So this 
transition should produce a MP between a pure hadronic and 
a pure quark phase. There are two ways by which one can construct the 
MP: the Maxwell construction (MC) and the Gibbs construction (GC). 
Below we consider both possibilities. 

The Gibbs conditions for the mixed phase are 
\begin{eqnarray}
P_1(\mu_B,\mu_Q) &=& P_2(\mu_B,\mu_Q)\,,\\ \label{eq:10}
\mu_B \,=\, \mu_{B1} &=& \mu_{B2}\,,\\
\mu_Q \,=\, \mu_{Q1} &=& \mu_{Q2}\,.  
\end{eqnarray}
Here and below $1$ stands for the hadronic phase and $2$ stands for 
the quark phase. 

It is well known 
that in the case of two chemical potentials the Gibbs conditions 
(10) - (12) can be fulfilled only if the coexisting 
phases have opposite electric charges and the condition of global neutrality
is imposed \cite{glend}. This condition can be written as
\be
{\bar \rho}_Q = (1-\lambda)\hsp\rho_{Q1}(\mu_B,\mu_Q)+
\lambda\hsp\rho_{Q2}(\mu_B,\mu_Q) = 0\,.
\ee 
Then the volume averaged energy density in the MP is calculated as
\begin{eqnarray}
{\bar \epsilon} &=&(1-\lambda)\hsp\epsilon_1(\mu_B,\mu_Q) + 
\lambda\hsp\epsilon_2(\mu_B,\mu_Q)\,,
\label{eq:12}
\end{eqnarray}
where $\lambda=V_2/V$ is the volume fraction of quark phase. 

Thus, the mixed phase is a very inhomogeneous state of matter 
with intermittent domains of two different phases. Therefore 
realistic approaches must take into account not only 
differences in the bulk properties of these phases, but also 
additional contributions to the thermodynamic potential 
coming from the interface energy and electrostatic energy 
associated with theses domains. First attempt to perform 
such calculations have been done in \cite{pethick} but 
due to significant uncertainties in the model parameters, as 
{\it e.g.} interface energy, the results are not conclusive 
yet. As pointed out in ref. \cite{pasta}, it may happen that 
the GC mixed phase is energetically too expensive and may be 
expelled from the star at all. Then the situation is closer 
to the MC case, where two pure phases are in direct contact with 
each other. 
This situation corresponds to the Maxwell construction of the 
mixed phase defined by the conditions : 
\begin{eqnarray}
P_1(\mu_B,\mu_Q) &=& P_2(\mu_B,\mu_Q) \\
\mu_B \,=\, \mu_{B1} &=& \mu_{B2}
\end{eqnarray}
They mean that the baryon chemical potential is continuous, but the 
electric chemical potential $\mu_Q$ jumps at the 
interface between the two phases. Also, contrary to the Gibbs 
construction, where the pressure in the mixed phase increases with baryon 
density, Maxwell construction corresponds to 
constant pressure in the density interval of the mixed 
phase.

\begin{figure}[htb]
\vskip 0.2in
\centerline{\includegraphics[width=3in]{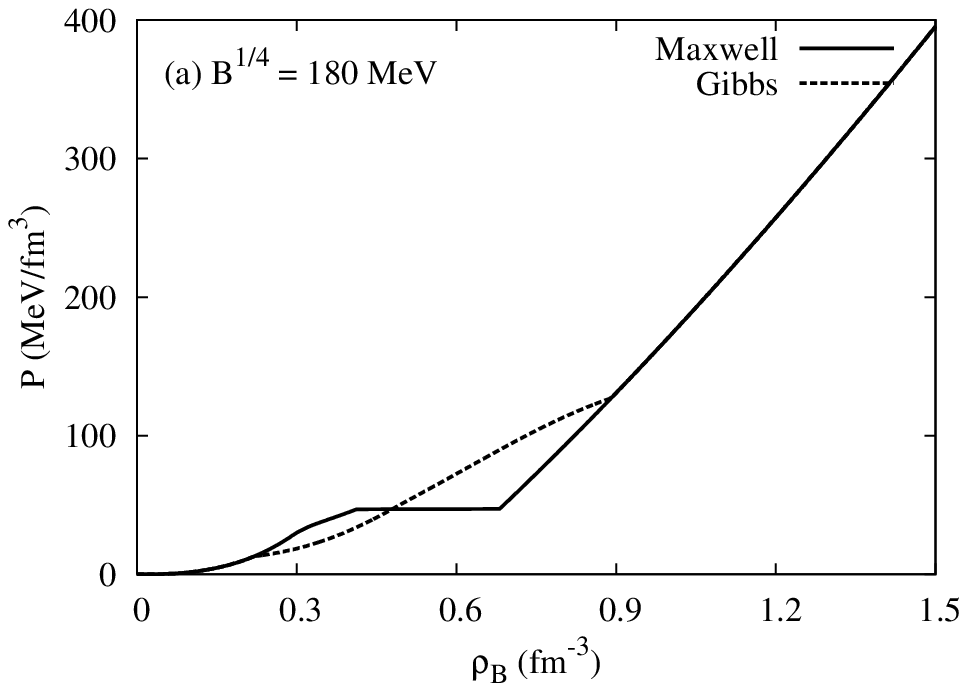}
\includegraphics[width=3in]{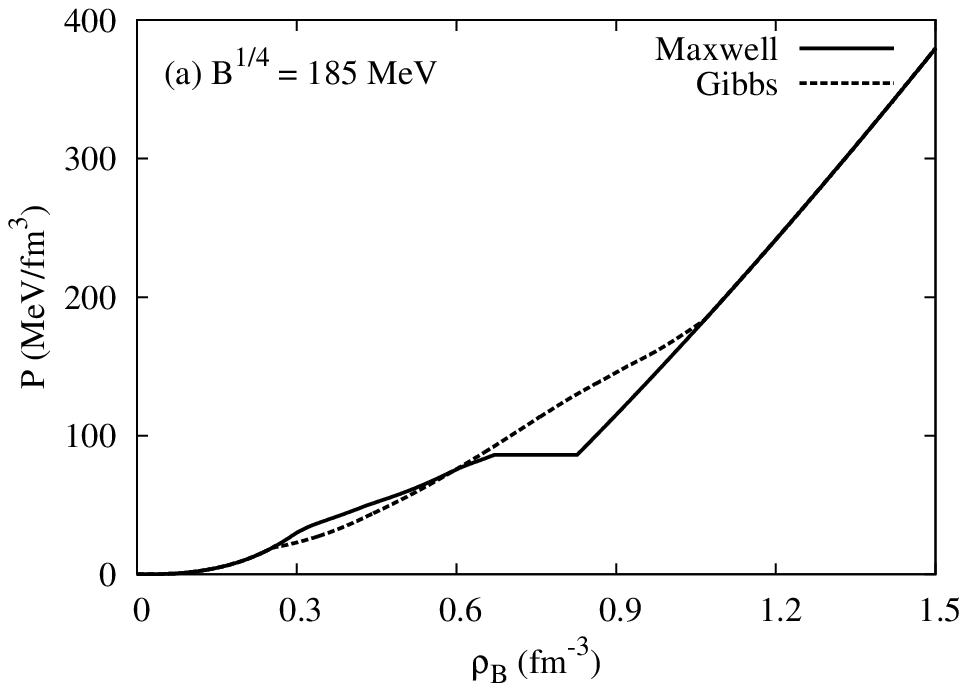}}
\caption{Equations of state for a) $\bg$ = 180 MeV and b) $\bg$ 
=185 MeV, for Maxwell and Gibbs construction of the mixed 
phase.} 
\end{figure}

Figures 1a and 1b show the equations of state obtained with GC 
and MC for two cases, $B^{1/4} = $ 180 MeV and 185 MeV, 
respectively. As can be seen from Fig. 1a, for MC the mixed 
phase starts at $\rb = 0.41 \, \dn$ and ends at $0.68 \, \dn$, whereas 
for the GC these values are $0.23 \, \dn$ and $0.89 \, \dn$ 
respectively. So in the case of GC the phase transition starts 
early and the width of the MP region is much boader compared 
to that in MC. As one increases the Bag constant B the width 
of the MP region further increases for GC, as can be seen in 
Fig. 1b. In this case the MP starts at $\rb = 0.26 \, \dn$ and 
ends at $1.06 \, \dn$. On the other hand, for the MC the width of 
the MP region decreases. For this case the MP starts at 
$\rb = 0.67 \, \dn$ and ends at $0.83 \, \dn$. These results will 
certainly affect the star properties which we will discuss 
in the next section. 

\begin{figure}[htb]
\vskip 0.2in
\centerline{\includegraphics[width=3in]{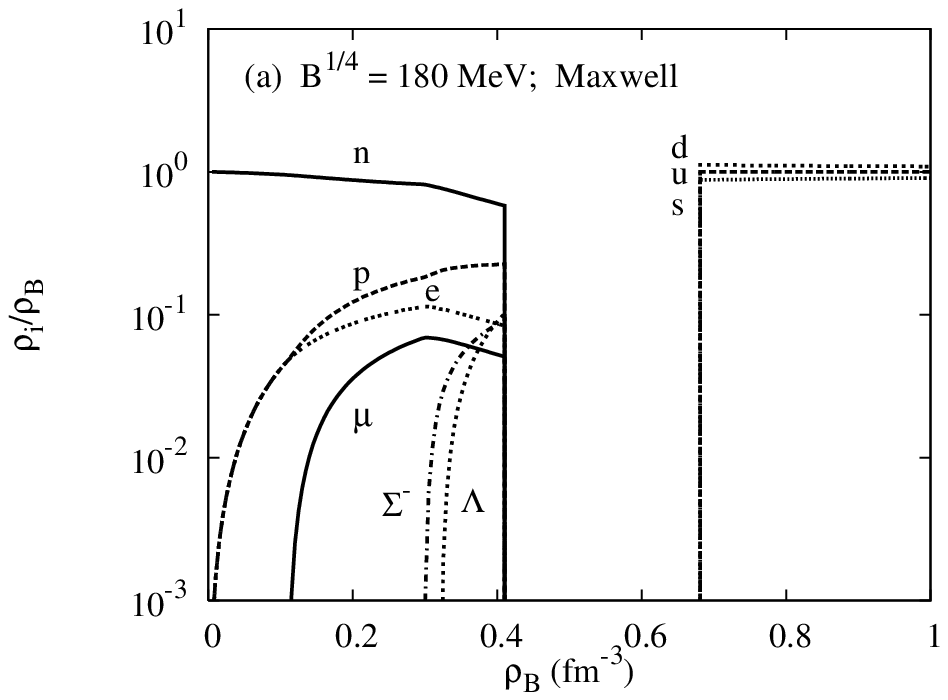}
\includegraphics[width=3in]{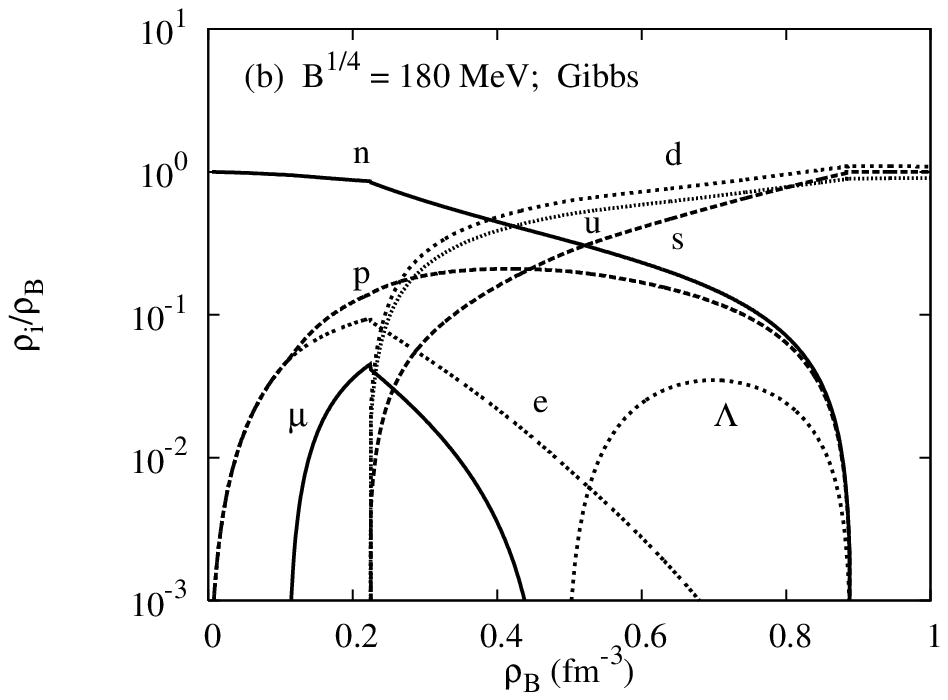}}
\centerline{\includegraphics[width=3in]{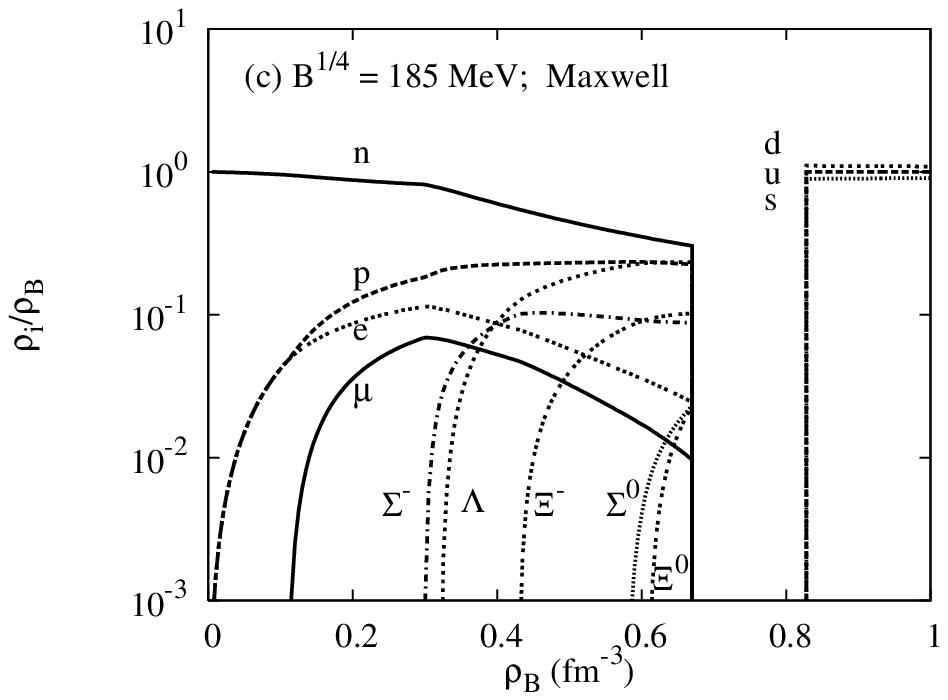}
\includegraphics[width=3in]{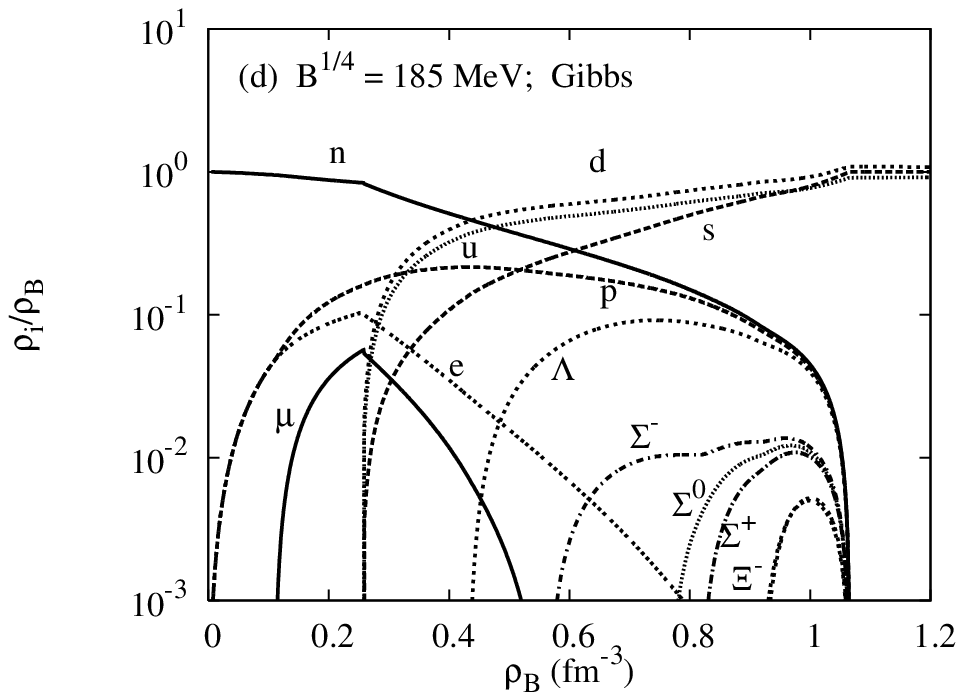}}
\caption{Particle compositions for a) $\bg$ = 180 MeV MC, 
b) $\bg$ = 180 MeV GC, c) $\bg$ = 185 MeV MC and 
d) $\bg$ = 185 MeV GC.}
\end{figure}

In Fig. 2 we have plotted the particle abundances for all the cases 
discussed above. In Figs. 2a and 2b the results are for $\bg = 180$ MeV,
 for the MC and GC, respectively. From Fig. 2b, 
{\it i.e.} for GC,  we see that only one hyperon, {\it i.e.} 
$\Lambda$, is present in the medium. On the other hand, for the case 
of MC (Fig. 2a) both $\Sigma^-$ and $\Lambda$ hyperons appear in the 
medium. This happens because in the case of GC the phase transition 
starts early, and as a result, hyperons can appear only in the MP. 
However due to the presence of the strange quark, hyperon production 
in the MP is suppressed and as a result $\Sigma^-$ does not appear 
at all. For MC the phase transition starts much later allowing  
the $\Sigma^-$ to appear in the hadronic phase. It is interesting 
that the case of $\bg = 185$ MeV exhibits a completely different 
picture as shown in Figs. 2c and 2d. Firstly, in the case of GC, 
the MP region is broader as compared with the case of $\bg = 180$ 
MeV. This allows almost all the hyperons, except $\Xi^0$, to be 
present in the matter, the $\Sigma^-$ appears after $\Lambda$. For 
the case of MC, as the MP begins at a higher density, almost all the
 hyperons are present in the hadronic phase. But, contrary to the 
case of GC, the $\Sigma^-$ appears before $\Lambda$. So the particle 
cocktail is rather different for the two constructions.

\section{Properties of compact stars}

Having obtained the EOSs and the particle abundances we now calculate 
the properties of compact stars with the deconfinement phase 
transition. We treat the matter 
to be an ideal fluid and obtain the star structure by solving the TOV 
equations with the corresponding EOS as an input (see details 
in \cite{gle}). 
The masses and radii of stars are calculated as a function of the 
central baryon density. Results of our calculations for MC and GC are 
shown in Figs.  
3a and 3b, for $\bg$ = 180 MeV and 185 MeV, respectively.

\begin{figure}[htb]
\vskip 0.2in
\centerline{\includegraphics[width=3in]{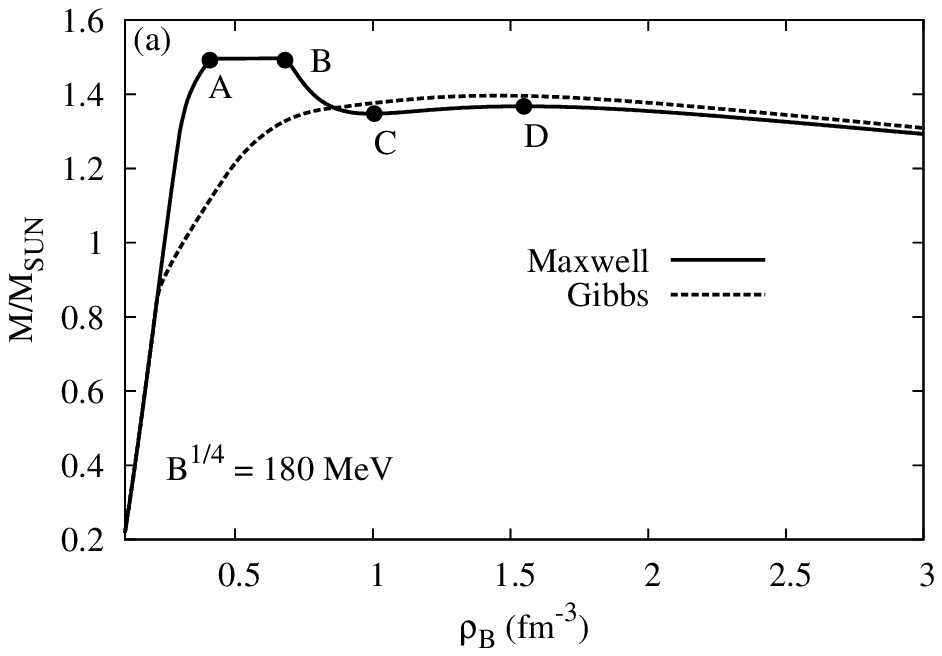}
\includegraphics[width=3in]{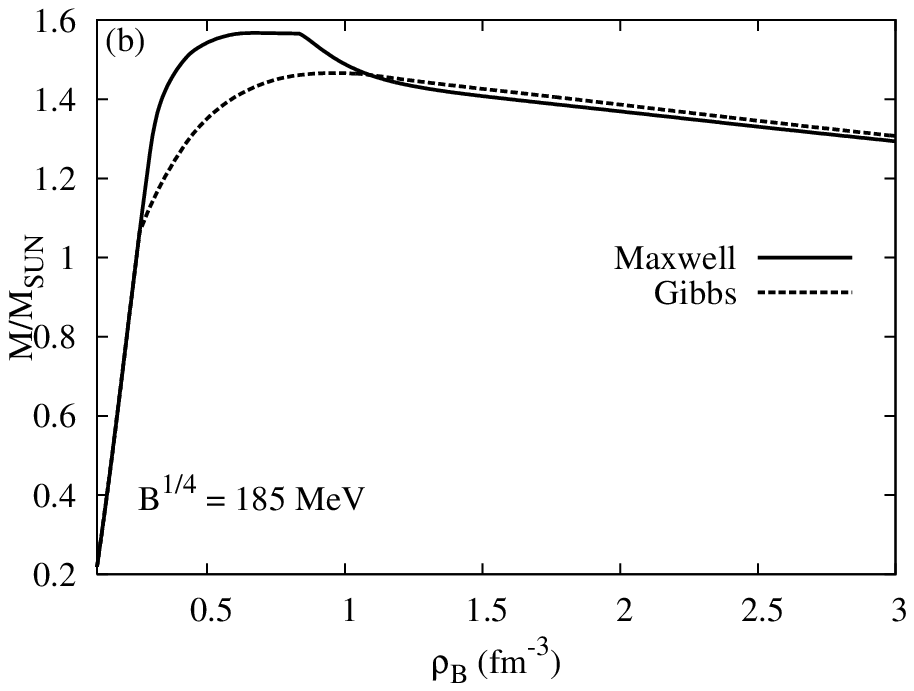}}
\caption{Mass of a star (in solar units) as a function of baryon 
density for a) $\bg$ = 180 MeV and b) $\bg$ =185 MeV. See text 
for further explanation.} 
\end{figure}

In Fig. 3a one can see that the maximum 
mass for the MC is noticeably higher than that for the GC case. The values 
are $1.493 M_\odot$ and $1.396 M_\odot$ respectively. This happens 
because in the MC the phase transition starts quite late compared to 
the GC, that allows the star to stay longer in the hadronic phase. 
Moreover, for the MC case there is a range of central baryon densities, 
from $0.41 \, \dn$ to $0.68 \, \dn$, corresponding to the MP, which are 
not allowed in the star.  
As a result there appears a plateau over this range of central baryon 
densities. As calculations show, stable star configurations are located 
on the left side of the plateau {\it i.e.} till the point A. The 
region from A to B is not accessible inside the star. On the right side of 
the plateau  there is a region, from B to C, where the mass 
decreases with increasing central baryon 
density, that corresponds to be the unstable configurations.
Another interesting feature of the MC stars is the appearance 
of the growing portion of the curve between C and D. This gives rise to 
a new family of stable stars (twin stars) 
around the mass of $1.34 M_\odot$ similar to the situation studied in 
ref. \cite{plb}. However there is no such a stable solution for GC. 
The main difference of twin stars from the normal neutron stars is 
that they contain a large quark matter core (see also Fig. 5a). 

For $\bg$ = 185 MeV the maximum masses are $1.567 M_\odot$ and 
$1.466 M_\odot$ for the MC and GC cases respectively. 
The forbidden range of baryon densities for the MC case is much smaller, 
ranging from $0.67 \, \dn$ to $0.83 \, \dn$ only. However, there is 
no stable twin star solution for this value of the Bag pressure. 
This situation can be explained from the well known fact that a significant 
quark core can appear only if the density jump is large enough 
\cite{migdal}.

\begin{figure}[htb]
\vskip 0.2in
\centerline{\includegraphics[width=3in]{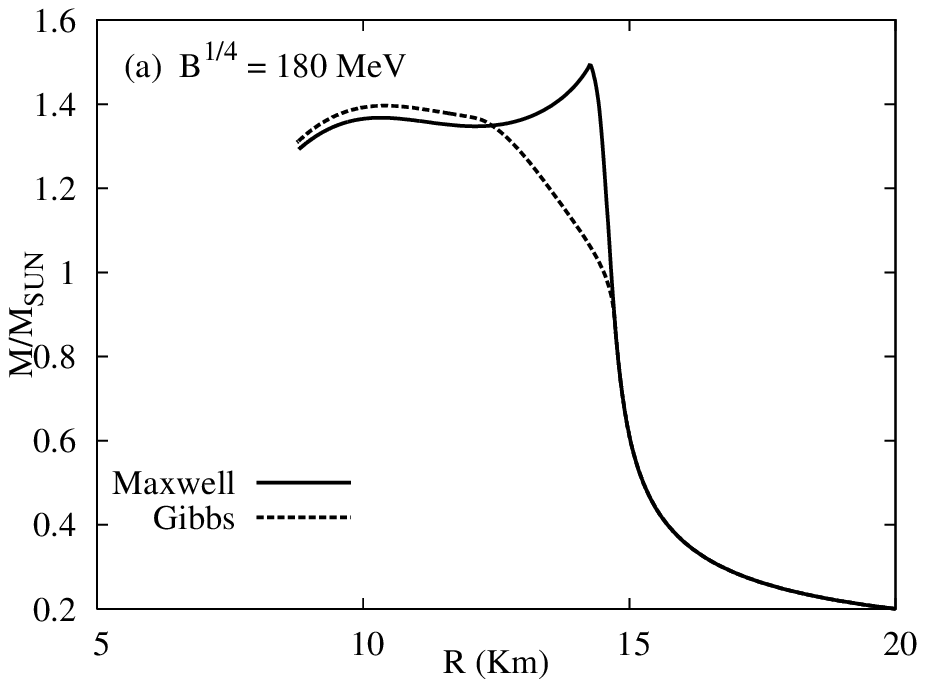}
\includegraphics[width=3in]{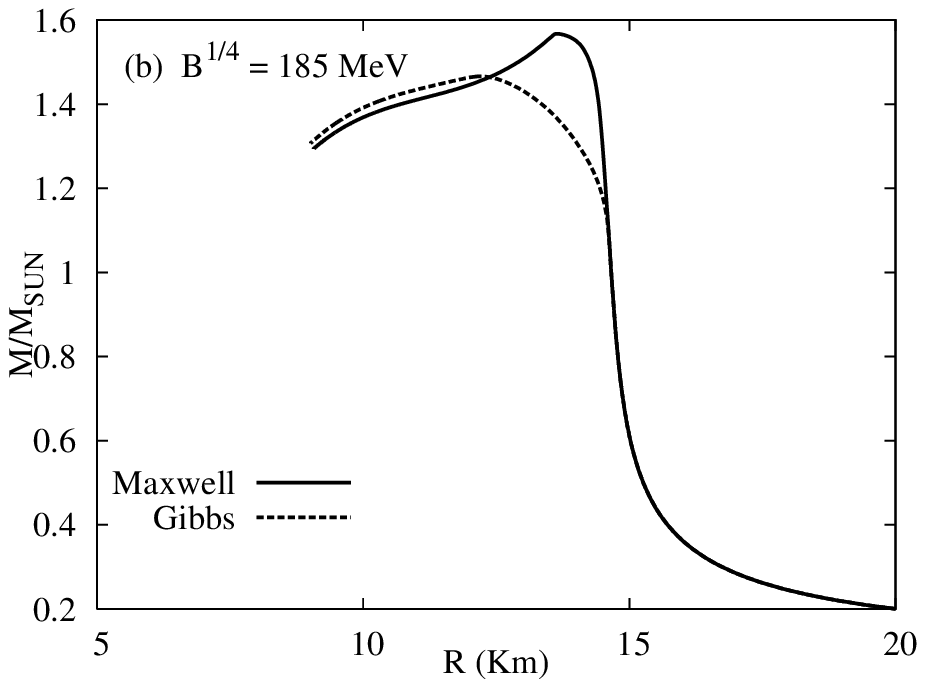}}
\caption{Mass-Radius relation of the stars for a) $\bg$ = 180 MeV 
and b) $\bg$ = 185 MeV. }
\end{figure}

The features discussed above are even more obvious if one looks at the 
mass-radius plots shown in Fig. 4. As one can see in Figs. 4a and 
4b the mass-radius relations are very different for MC and 
GC stars. Furthermore, the appearance of two maxima associated with the 
twin stars is very obvious in Fig. 4a. The cusps in the MC curves 
correspond to the plateau regions of Figs. 3a,b.

We have calculated also the baryon density profiles of MC and GC stars 
for the same central density. They are given in Figs. 5a and 5b. Figure 
5a shows a sudden jump in the baryon density for the MC case whereas in 
the GC case the profile is smooth. As Fig. 5b shows there is no jump 
for the case $\bg$ = 185 MeV and there is no stable star with a quark 
core. 

\begin{figure}[htb]
\vskip 0.2in
\centerline{\includegraphics[width=3in]{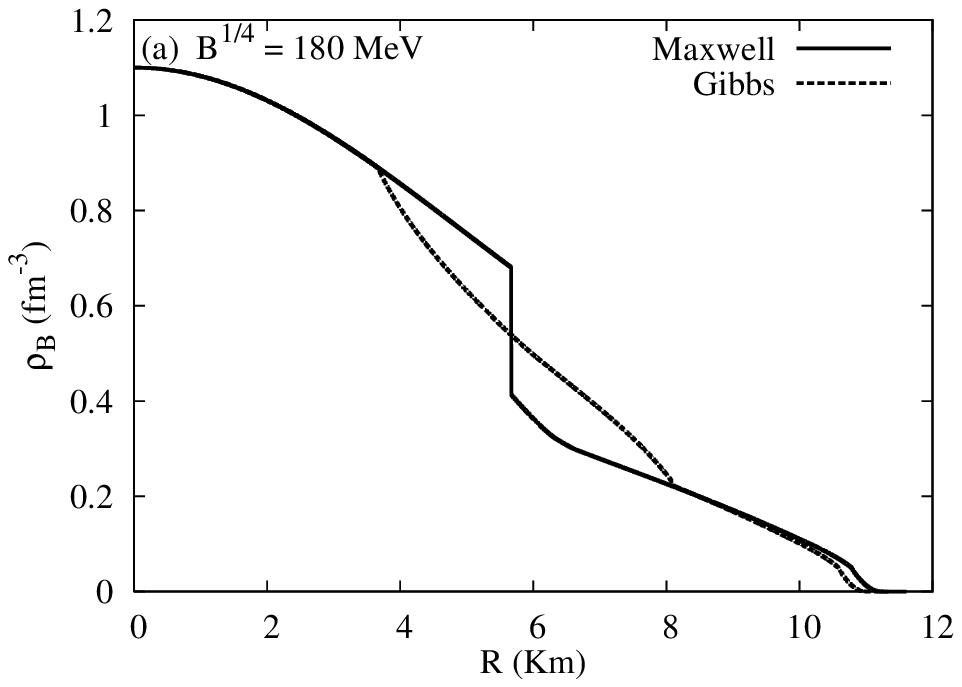}
{\includegraphics[width=3in]{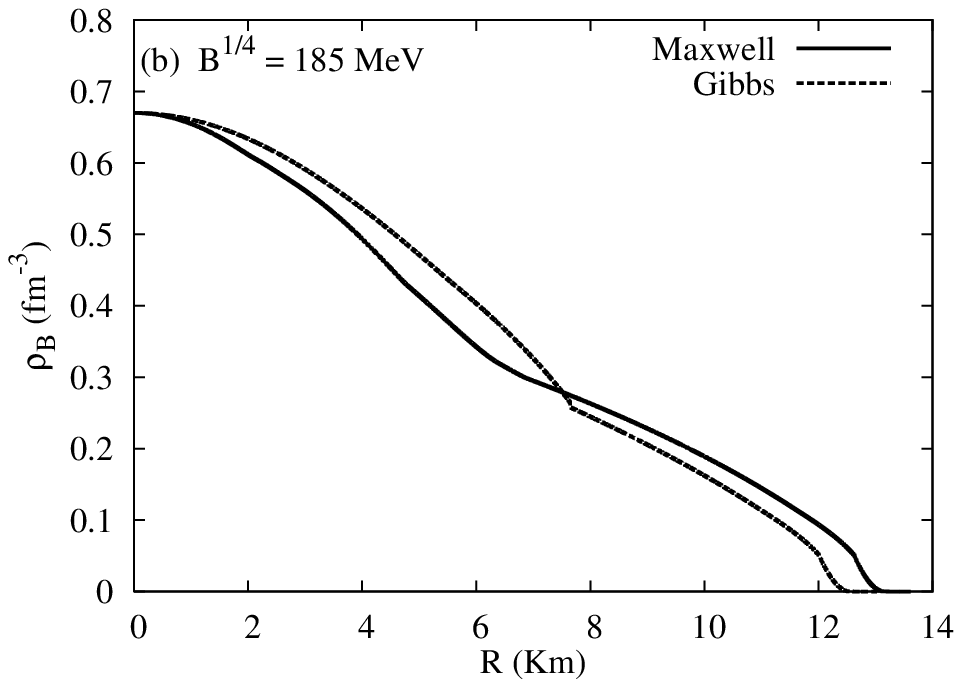}}}
\caption{Comparison of baryon density profiles for MC and GC stars 
calculated for a) $\bg$ = 180 MeV and b) $\bg$ = 185 MeV. }
\end{figure}
\section{Jump in electron chemical potential}

In this section we discuss possible implications of the jump in the 
baryon density of the MC stars as illustrated in Fig. 5a. Figure 6 
shows the radial profiles of the electron chemical potential 
$\mu_e$ for the 
case of $\bg$ = 180 MeV. As expected, in the GC case $\mu_e$ 
evolves smoothly over the mixed phase reaching almost zero in the 
pure quark phase. But in the MC case $\mu_e$ has a jump at the transition 
between hadronic and quark phases. In the quark core $\mu_e$ is 
very low about 10 MeV and is almost constant, but in the hadronic 
phase $\mu_e$ jumps to a high value, about 196 MeV. Then at larger 
radii $\mu_e$ evolves slowly and finally decrease to small values 
as one approaches the crust. 

This is an interesting situation in the sense that such a 
discontinuity in the chemical potential of electrons would lead 
to the flow of electrons across the discontinuity surface from the 
region with higher $\mu_e$ to the region with lower $\mu_e$ 
inside the star. This flow will be terminated by the electric 
field generated due to the charge separation. The equilibrium 
condition can be expressed as 
\begin{equation}
\mu_{e1} - e \Phi_1 = \mu_{e2} - e\Phi_2
\end{equation}
where $\mu_{e1}$ and $\mu_{e2}$ are the electron chemical 
potentials in the hadronic and quark phases, and  $\Phi_{1,2}$ 
are the electrostatic potentials far away from the discontinuity. 

\begin{figure}[h]
\vskip 0.2in
\centerline{\includegraphics[width=3.5in]{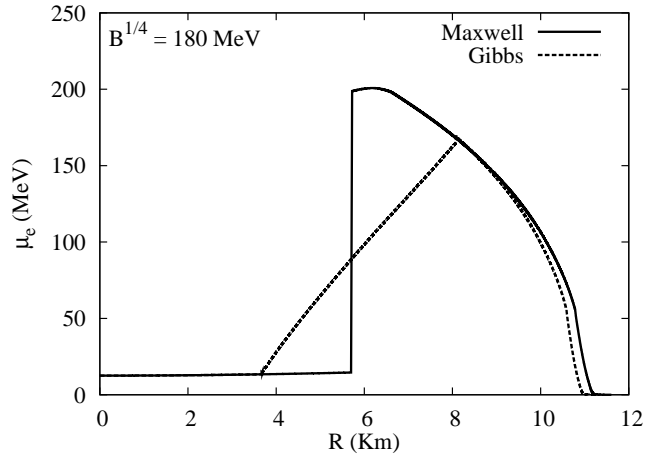}}
\caption{Profiles of electron chemical potential for MC and GC stars 
for $\bg$ = 180 MeV. }
\end{figure}

The profile of the electrostatic potential $\Phi(r)$ over the 
discontinuity surface can be found from the Poisson equation in 
combination with the Thomas-Fermi approximation for the electron density
$\rho_e(r)= {{k_F^3(r)} \over {3\pi^2}}$ where $k_F(r)$ is the 
local Fermi momentum of the electrons (see 
details in ref.\cite{ebel}). In the ultrarelativistic limit $\mu_1 
>> m_e$, $\mu_2 >> m_e$, the result can be obtained in the 
analytic form and the maximum electric field is expressed as 
\begin{equation}
E_0 = {{\mu_{e1}\mu_{e2}} \over {e \pi}}\times {{\mu_{e1}-\mu_{e2}} 
\over {\mu_{e1}+\mu_{e2}}}
\end{equation}
For the particular case shown in Fig. 6 we have $\mu_{e1}$ = 10 MeV 
and $\mu_{e2}$ = 196 MeV, that gives $E_0 \approx 3$ MV/fm.  
This is a very strong field, about 1000 times the critical field 
needed for the spontaneous electron-positron pair production in 
vacuum. In fact, this pair 
production process is Pauli blocked in the considered case. 
Nevertheless, such a strong electric field may lead to interesting 
phenomena such as e.g. generation of strong magnetic field in the 
rotating stars. The generation of strong electric fields at the 
bare boundary of a quark star was first discussed in ref. \cite{alcock}.

\section{Summary}

To summarise, we have compared the Gibbs and Maxwell constructions 
of the mixed phase in the context of deconfinement phase 
transition in compact stars. For this purpose we have used a RMF 
model (TM1YY) for the hadronic phase and the MIT Bag model for the 
quark phase. We have found that the EOSs are very different in 
these two cases: the MP region occupies a much broader density 
interval for GC as compared with the MC one. As we increase the 
Bag constant the width of the MP region in case of GC increases 
whereas it reduces for the MC case. The particle compositions are 
also found to be very different for the two cases. We then use 
these EOSs to calculate the star characteristics. The maximum 
mass is found to be different for the MC and GC cases. 
Furthermore, for $\bg$=180 MeV a stable solution corresponding 
to second family of compact stars is obtained for MC. The baryon 
density profiles show a sharp jump for the MC case.

We have also studied the behaviour of the electron chemical 
potential $\mu_e$ across the star and found that it jumps sharply, 
for MC, at the phase transition boundary. 
The jump is about 185 MeV for $\bg$ = 180 
MeV. We point out that this jump will lead to redistribution of electrons 
and generation of a strong electric field at the phase transition boundary.
We are planning to study this interesting effect in the future. 

\section*{Acknowledgements}
The authors thank L. M. Satarov for fruitful discussions. The work of 
AB was partially supported by Alexander von Humboldt Foundation, 
Council of Scientific and Industrial Research (India) and UGC (UPE \& 
DRS grant). AB also thanks FIAS for the kind hospitality. This work was 
partly supported by the DFG grant 436 RUS113/711/0-2 and the grants 
RFFI 09-02-91331, NS-3004.2008.2 (Russia).

\end{document}